# The Binary Zoo: The Calculation of Production Rates of Binaries Through 2+1 Encounters in Globular Clusters

Melvyn B. Davies *

*Theoretical Astrophysics, Caltech, Pasadena CA 91125.*

**ABSTRACT**
In studying encounters between binaries and single stars, one is interested in three classes of events: exchanges of stars, hardening of the original binary by a third star, and the production of merged objects. We present a means for computing cross sections for these three outcomes for an arbitrary binary and single star as might be found in the core of a globular cluster. The cross sections for a number of binaries in various stellar populations are then computed. We consider multiple encounters and the ultimate fate of a population of binaries fed into the cores of different globular cluster models. We see that the presence of only a relatively small number of binaries (containing 10% of the stars) will boost the production rate of astrophysically-interesting objects by a factor of at least a few over the rates expected from encounters between single stars. In particular, the ratio of smothered neutron stars to low-mass X-ray binaries (LMXBs) may be greatly increased, possibly explaining, in part, the excess of millisecond pulsars compared to LMXBs.

**Key words:** stellar: evolution – globular clusters

## INTRODUCTION

Encounters involving binaries may be important in globular clusters even if their relative fraction is low, owing to their large cross section. When considering the evolution of dense stellar systems, and stellar encounters within them, one would therefore wish to account for encounters involving binaries, which may provide the energy source to support a core from collapse and may produce many of the stellar exotica seen; including blue stragglers, low-mass X-ray binaries (LMXBs), and millisecond pulsars (MSPs).

Encounters between binaries and single stars treating the stars as point masses have been considered in numerous works (Hut 1994, and references contained therein). Some allowance has also been made for the finite size of the stars in three-body simulations (see for example, Sigurdsson & Phinney 1994). Cleary & Monaghan (1990), and Davies, Benz & Hills (1993, 1994) also considered the hydrodynamical effects by performing some encounters using a smoothed particle hydrodynamics (SPH) code.

Studies of encounters between binaries and single stars show three main outcomes: the incoming star replaces one of the original components forming a new, detached, binary in a so-called *clean exchange*, the incoming star simply hardens the binary without any exchange occuring – a so-called *fly-by*, or a merger between two of the stars occurs. One would wish to have expressions for these three outcomes; the exchange cross section, $\sigma_{cx}$; the fly-by cross section, $\sigma_{fb}$; and the cross section to form a merged object, $\sigma_{mb}$. The last outcome bifurcates; the merged object may be bound to the third star, producing a so-called *merged binary* (MB), or it may be unbound, an event that has been dubbed a *scattering-induced merger* (SIM).

---

* Current Address: Institute of Astronomy, Madingley Road, Cambridge CB3 0HA

We seek a means for computing these cross sections for various stellar masses, binary separations, and relative velocities. Rather than consider the three stellar masses, we can consider two mass ratios; the ratio of the secondary to the primary in the original binary, $q_1$, and the ratio of the incoming field star to that of the primary mass, $q_2$. We performed a series of three-body encounters for a range of mass ratios, $q_1$, $q_2$, and from them derived a number of parameters used to calculate the various cross sections. We then apply these cross sections, and feed various binaries into various stellar populations and compute the distribution of objects produced.

The parameters required to calculate the various cross sections are discussed in §2. The results of our three-body runs are given in §3. We apply these results in §4 and compute the various cross sections for various binaries in various stellar populations. We consider multiple encounters and the ultimate fate of binaries fed into particular globular clusters in §5.

**Table 1.** Parameters obtained for encounters involving circular binaries as a function of the ratio of the mass of the primary to the secondary in the original binary ($q_1$) and the ratio of field-star mass to the mass of the primary ($q_2$). See text for definitions of the parameters.

| $q_1$ | $q_2$ | $k_{ex}$ | $k_{fb}$ | $k_{12}$ | $k_{13}$ | $k_{23}$ | $\gamma_{12}$ | $\gamma_{13}$ | $\gamma_{23}$ |
|---|---|---|---|---|---|---|---|---|---|
| 0.312 | 2.000 | 45832 | 13747 | 36150 | 48099 | 29767 | 0.850 | 1.060 | 0.822 |
| 0.312 | 1.500 | 33231 | 10137 | 21193 | 22754 | 32566 | 0.772 | 0.880 | 0.967 |
| 0.312 | 1.000 | 17141 | 8108 | 24370 | 11412 | 24467 | 0.908 | 0.720 | 0.961 |
| 0.312 | 0.571 | 8151 | 4051 | 9265 | 7244 | 9335 | 0.626 | 0.564 | 0.708 |
| 0.571 | 2.000 | 31351 | 9517 | 27286 | 18589 | 23694 | 0.758 | 0.627 | 0.750 |
| 0.571 | 1.500 | 16849 | 3874 | 17623 | 14390 | 14373 | 0.756 | 0.626 | 0.665 |
| 0.571 | 1.000 | 11648 | 5354 | 12432 | 7600 | 12281 | 0.625 | 0.417 | 0.616 |
| 0.571 | 0.571 | 3187 | 4324 | 4965 | 4148 | 5147 | 0.451 | 0.412 | 0.540 |
| 1.000 | 2.000 | 19281 | 7853 | 16210 | 13000 | 13662 | 0.687 | 0.550 | 0.568 |
| 1.000 | 1.500 | 13104 | 5006 | 10941 | 10246 | 11370 | 0.566 | 0.531 | 0.594 |
| 1.000 | 1.000 | 6222 | 4883 | 5848 | 5030 | 5532 | 0.436 | 0.378 | 0.405 |
| 1.000 | 0.570 | 813 | 3370 | 3020 | 2360 | 2748 | 0.432 | 0.365 | 0.429 |

**Table 2.** Parameters obtained for encounters involving circular binaries as a function of the ratio of the mass of the primary to the secondary in the original binary ($q_1$) and the ratio of field-star mass to the mass of the primary ($q_2$). See text for definitions of the parameters.

| $q_1$ | $q_2$ | $k_{12x}$ | $k_{13x}$ | $k_{23x}$ | $\gamma_{12x}$ | $\gamma_{13x}$ | $\gamma_{23x}$ | $k_{12f}$ | $k_{13f}$ | $k_{23f}$ | $\gamma_{12f}$ | $\gamma_{13f}$ | $\gamma_{23f}$ |
|---|---|---|---|---|---|---|---|---|---|---|---|---|---|
| 0.312 | 2.000 | 12627 | 47205 | 29767 | 0.776 | 1.055 | 0.822 | 23944 | 0 | 0 | 1.017 | 0.000 | 0.000 |
| 0.312 | 1.500 | 8174 | 21812 | 32385 | 0.608 | 0.870 | 0.965 | 28975 | 0 | 0 | 1.324 | 0.000 | 0.000 |
| 0.312 | 1.000 | 9135 | 11078 | 23090 | 0.739 | 0.714 | 0.947 | 34157 | 0 | 0 | 1.483 | 0.000 | 0.000 |
| 0.312 | 0.571 | 4861 | 6814 | 7767 | 0.553 | 0.575 | 0.692 | 2544 | 302 | 431 | 0.699 | 0.433 | 0.745 |
| 0.571 | 2.000 | 3326 | 18563 | 23694 | 0.678 | 0.628 | 0.750 | 12520 | 0 | 0 | 0.921 | 0.000 | 0.000 |
| 0.571 | 1.500 | 8742 | 14061 | 13812 | 0.698 | 0.625 | 0.656 | 6901 | 456 | 0 | 0.788 | 0.833 | 0.000 |
| 0.571 | 1.000 | 5650 | 7382 | 11851 | 0.528 | 0.431 | 0.630 | 6109 | 212 | 183 | 0.836 | 0.228 | 0.319 |
| 0.571 | 0.571 | 1170 | 2472 | 2943 | 0.303 | 0.412 | 0.533 | 2959 | 1129 | 1407 | 0.555 | 0.432 | 0.555 |
| 1.000 | 2.000 | 6930 | 12892 | 14167 | 0.590 | 0.557 | 0.591 | 10785 | 78 | 33 | 0.912 | 0.446 | 0.000 |
| 1.000 | 1.500 | 4251 | 10224 | 10313 | 0.435 | 0.561 | 0.588 | 5972 | 154 | 939 | 0.767 | 0.178 | 0.795 |
| 1.000 | 1.000 | 2377 | 3771 | 4662 | 0.351 | 0.386 | 0.455 | 3238 | 681 | 569 | 0.684 | 0.340 | 0.284 |
| 1.000 | 0.570 | 242 | 581 | 847 | 0.317 | 0.345 | 0.469 | 1534 | 1227 | 1517 | 0.456 | 0.461 | 0.564 |

## 2 THEORETICAL CROSS SECTIONS AND SCALING LAWS

The cross section for a single star to pass within a distance $R_{min}$ of the center of mass of a binary is given by $\sigma = \pi R_{min}^2 (1 + V_c^2/V_\infty^2)$, where $V_\infty$ is the relative speed at infinity, and $V_c$ is the relative speed at which the system has zero total energy and is given by $V_c^2 = F(M_1, M_2, M_3)/d = GM_1 M_2(M_1 + M_2 + M_3)/M_3(M_1 + M_2)d$, where $M_1$ is the mass of the primary, $M_2$ the mass of the secondary and $M_3$ is the mass of the incoming, single star. In order for an exchange encounter to occur where the impacting star replaces one of the binary components, one might expect $R_{min} \sim d$, where $d$ is the separation of the two components in the original binary. For a binary of separation $\sim 10$ AU, $V_c \simeq 10$ km/s (a typical value for the velocity dispersion in a globular cluster). Hence for hard binaries in globular clusters (where $d \sim 0.1 - 0.5$ AU), $V_c \gg V_\infty$, and the exchange cross section can be written as

$$\sigma_{ex} = k_{ex}(q_1, q_2) \pi d^2 \cdot \frac{GM_1 F(q_1, q_2)}{d} \cdot \frac{1}{V_\infty^2} \quad (1)$$

where $q_1 = M_2/M_1$, $q_2 = M_3/M_1$, and $F(q_1, q_2) = q_1(1 + q_1 + q_2)/q_2(1 + q_1)$. The constant $k_{ex}(q_1, q_2)$ has to be determined through numerical simulations.

Similarly, one may write an expression for the cross section for fly-by encounters, where the components of the binary remain unchanged, but the binary is hardened by some minimum amount (taken to be 10% in our later calculations here).

$$\sigma_{fb} = k_{fb}(q_1, q_2) \pi d^2 \cdot \frac{GM_1 F(q_1, q_2)}{d} \cdot \frac{1}{V_\infty^2} \quad (2)$$

where again $k_{fb}(q_1, q_2)$ has to be determined through numerical simulations.

One may also compute the cross section for two of the three stars to pass within some minimum distance during an encounter. Hut and Inagaki (1985) found that such a cross section can be written in the following form, for the $i$th and $j$th stars,

$$\sigma_{rmin}(i, j) = k_{rmin}(q_1, q_2, i, j) \pi d^2 \frac{GM_1 F(q_1, q_2)}{d} \cdot \frac{1}{V_\infty^2}$$

**Table 3.** Parameters obtained for encounters involving eccentric binaries as a function of the ratio of the mass of the primary to the secondary in the original binary ($q_1$) and the ratio of field-star mass to the mass of the primary ($q_2$). See text for definitions of the parameters.

| $q_1$ | $q_2$ | $k_{ex}$ | $k_{fb}$ | $k_{12}$ | $k_{13}$ | $k_{23}$ | $\gamma_{12}$ | $\gamma_{13}$ | $\gamma_{23}$ |
|---|---|---|---|---|---|---|---|---|---|
| 0.312 | 2.000 | 60643 | 15779 | 92862 | 31404 | 48135 | 0.906 | 0.799 | 0.832 |
| 0.312 | 1.500 | 38922 | 12143 | 52266 | 21366 | 25308 | 0.794 | 0.836 | 0.822 |
| 0.312 | 1.000 | 22535 | 9135 | 52055 | 18554 | 19072 | 0.897 | 0.823 | 0.725 |
| 0.312 | 0.571 | 9399 | 4711 | 13774 | 6189 | 7142 | 0.616 | 0.491 | 0.556 |
| 0.571 | 2.000 | 41067 | 10937 | 51170 | 25125 | 25297 | 0.720 | 0.672 | 0.643 |
| 0.571 | 1.500 | 26443 | 8517 | 32028 | 14190 | 19671 | 0.634 | 0.538 | 0.598 |
| 0.571 | 1.000 | 13536 | 5783 | 20371 | 11237 | 10781 | 0.619 | 0.518 | 0.521 |
| 0.571 | 0.571 | 4314 | 4656 | 8133 | 5003 | 5569 | 0.470 | 0.394 | 0.436 |
| 1.000 | 2.000 | 25546 | 8202 | 37341 | 15466 | 14579 | 0.699 | 0.523 | 0.526 |
| 1.000 | 1.500 | 16429 | 7539 | 22266 | 11558 | 12012 | 0.600 | 0.487 | 0.517 |
| 1.000 | 1.000 | 7328 | 5078 | 11315 | 7901 | 6093 | 0.501 | 0.461 | 0.364 |
| 1.000 | 0.571 | 1187 | 3896 | 4882 | 3184 | 3779 | 0.453 | 0.345 | 0.424 |

**Table 4.** Parameters obtained for encounters involving eccentric binaries as a function of the ratio of the mass of the primary to the secondary in the original binary ($q_1$) and the ratio of field-star mass to the mass of the primary ($q_2$). See text for definitions of the parameters.

| $q_1$ | $q_2$ | $k_{12x}$ | $k_{13x}$ | $k_{23x}$ | $\gamma_{12x}$ | $\gamma_{13x}$ | $\gamma_{23x}$ | $k_{12f}$ | $k_{13f}$ | $k_{23f}$ | $\gamma_{12f}$ | $\gamma_{13f}$ | $\gamma_{23f}$ |
|---|---|---|---|---|---|---|---|---|---|---|---|---|---|
| 0.312 | 2.000 | 39434 | 30488 | 49184 | 0.886 | 0.792 | 0.842 | 27707 | 0 | 0 | 0.855 | 0.000 | 0.000 |
| 0.312 | 1.500 | 27445 | 20883 | 24989 | 0.804 | 0.830 | 0.821 | 15966 | 0 | 0 | 0.720 | 0.000 | 0.000 |
| 0.312 | 1.000 | 24017 | 18057 | 19167 | 0.863 | 0.818 | 0.733 | 25629 | 0 | 0 | 1.092 | 0.000 | 0.000 |
| 0.312 | 0.571 | 6850 | 5283 | 6090 | 0.546 | 0.472 | 0.543 | 4546 | 481 | 1501 | 0.768 | 0.542 | 1.231 |
| 0.571 | 2.000 | 26972 | 24553 | 24917 | 0.679 | 0.668 | 0.641 | 14997 | 393 | 0 | 0.743 | 0.963 | 0.000 |
| 0.571 | 1.500 | 18230 | 13631 | 18970 | 0.609 | 0.533 | 0.601 | 12104 | 118 | 369 | 0.754 | 0.267 | 0.468 |
| 0.571 | 1.000 | 9125 | 10558 | 10262 | 0.526 | 0.530 | 0.543 | 8378 | 388 | 269 | 0.827 | 0.396 | 0.350 |
| 0.571 | 0.571 | 2394 | 2976 | 3223 | 0.396 | 0.384 | 0.429 | 4031 | 1359 | 1396 | 0.600 | 0.432 | 0.480 |
| 1.000 | 2.000 | 18656 | 15072 | 14023 | 0.627 | 0.525 | 0.525 | 11840 | 176 | 184 | 0.799 | 0.506 | 0.338 |
| 1.000 | 1.500 | 8246 | 10890 | 10846 | 0.478 | 0.493 | 0.512 | 11297 | 799 | 378 | 0.800 | 0.679 | 0.450 |
| 1.000 | 1.000 | 3337 | 6357 | 4312 | 0.389 | 0.490 | 0.360 | 4849 | 1277 | 919 | 0.600 | 0.459 | 0.319 |
| 1.000 | 0.571 | 506 | 900 | 1244 | 0.379 | 0.310 | 0.457 | 2854 | 1360 | 1455 | 0.514 | 0.396 | 0.416 |

$$\times \left( \frac{r_{\min}(i,j)}{d} \right)^{\gamma(i,j)} \quad (3)$$

where both $k_{\rm rmin}(q_1, q_2, i, j)$ and $\gamma(i,j)$ are to be determined (typically $\gamma(i,j) \sim 0.5$). We must find these constants for each of the three pairs; 1, 2; 1, 3; and 2, 3. We consider the three pairs separately to maintain generality; if two of the three stars are compact, we will not wish to consider close encounters between them (unless the binary separation is extremely small). We must also consider separately the cross section for encounters that contribute to either the exchange or fly-by cross section but which really led to some form of merger event. Again, we must consider each pair of stars separately. We thus have to compute *nine* values of $k_{\rm rmin}$ and $\gamma$.

To compute cross sections for mergers, we have to provide a value for the minimum distance, $r_{\min}(i,j)$, between the $i$th and $j$th stars. Here, we use the simple formula obtained by Kochanek (1992)

$$\frac{R_{\rm merg}}{R_i} = 1.7 \left( \frac{M_i + M_j}{2 M_i} \right)^{1/3} \quad (4)$$

where, for a given pair, the larger of the two values of $R_{\rm merg}$ is used.

We also have to differentiate between merger events leading to so-called merged binaries (MBs), where the merged object remains bound to the third star, and scattering induced mergers (SIMs), where the two objects are unbound. We may determine this branching ratio empirically for each set of simulations, thus having

$$\begin{aligned}\sigma_{\rm sim} &= \alpha_{\rm sim} \sigma_{\rm rmin} \\ \sigma_{\rm mb} &= (1 - \alpha_{\rm sim}) \sigma_{\rm rmin}\end{aligned} \quad (5)$$

where we have to compute an $\alpha$ for each of the three pairs of stars. We compute the cross sections in equation (4) for each of the three pairs and sum them (assuming all pairs contain no more than one compact star).

We compute the cross sections for *clean* exchanges by subtracting the cross section of exchanges that would really have led to merger events from the exchange cross section given in equation(1). We thus have

$$\sigma_{\rm cx} = \sigma_{\rm ex} - \sigma_{\rm rmin|ex} \quad (6)$$

where $\sigma_{\rm rmin|ex}$ is the cross section of merger events that were eroneously labelled as exchanges. Written out in full, $\sigma_{\rm rmin|ex} = i_{12}\sigma_{\rm rmin|ex}(1,2) + i_{13}\sigma_{\rm rmin|ex}(1,3) +$

$i_{23}\sigma_{\text{rmin}|\text{ex}}(2,3)$, where $i_{12}$, $i_{13}$, and $i_{23}$ are unity unless both stars in the pair are compact objects, in which case the value is zero. We apply a similar procedure to the fly-by cross section, with the cross section for clean fly-bys being given by

$$\sigma_{\text{cfb}} = \sigma_{\text{fb}} - \sigma_{\text{rmin}|\text{fb}} \quad (7)$$

where $\sigma_{\text{rmin}|\text{fb}} = i_{12}\sigma_{\text{rmin}|\text{fb}}(1,2) + i_{13}\sigma_{\text{rmin}|\text{fb}}(1,3) + i_{23}\sigma_{\text{rmin}|\text{fb}}(2,3)$, and, as before, $i_{12}$, $i_{13}$, and $i_{23}$ are unity unless both stars in the pair are compact objects, in which case the value is zero.

## 3 RESULTS

We performed simulations of three-body encounters for three different values of $q_1$ and four different values of $q_2$, thus yielding twelve different combinations. We considered both circular and eccentric (e=0.5) binaries. In all cases the binary had a separation, $d = 60R_\odot$, and $V_\infty = 10$km/s. Some of these sets of encounters have been discussed elsewhere (Davies, Benz & Hills [1993, 1994]).

For each set of encounters, we computed the various cross sections as described in the previous section, and thus calculated the parameters required to obtain the cross sections for a different binary/single star encounter. These parameters are listed in Tables 1-4, the units for the masses and separations of the stars being solar, with velocities being given in units of 10km/s.

### 3.1 Cross Sections for Particular Binaries

We first apply the results of our three-body runs to compute the cross sections for particular binaries for a range of incoming field stars. For simplicity here, we consider only binaries containing two $0.7M_\odot$ stars (i.e. $q_1 = 1.0$), and consider only binaries having an eccentricity, $e = 0.5$. In practice, the eccentricity of the original binaries has very little effect on the relative cross sections, rather all cross sections are increased by $\sim 50\%$ when comparing encounters involving circular binaries to those involving binaries of eccentricity, $e = 0.5$. We consider binaries of three different separations; $50R_\odot$, $100R_\odot$, and $200R_\odot$. In the following discussion, we will assume equipartition of kinetic energy, i.e. $M_i V_i^2 = $ a constant. This assumption will certainly be valid in the cores of clusters, where the relaxation time is short. The mean relative speed at infinity between two stellar classes 1 and 2 can be estimated by (Verbunt & Meylan 1988)

$$\left\langle \frac{1}{V_\infty} \right\rangle = \sqrt{\frac{6}{\pi} \frac{1}{V_1^2 + V_2^2}} \quad (8)$$

In Figure 1 we show the computed cross sections for mergers, clean exchanges, and clean fly-bys (shaded black, dark grey, and light grey respectively), as a function of the mass of the incoming field star, $M_3$. The cross sections are given in units of the clean exchange cross section obtained for the binary of separation, $d = 50R_\odot$, with a incoming field star of mass, $M_3 = 0.7M_\odot$. It is clear from this figure that the clean exchange cross section is a strong function of $M_3$; essentially no clean exchanges will occur when $M_3 \le 0.4M_\odot$. It should also be noted from this figure that the cross section for mergers is relatively independent of $M_3$. This result has been reported previously (Davies, Benz and Hills 1994). The total cross section for some interaction (i.e. $\sigma_{\text{cfb}} + \sigma_{\text{cx}} + \sigma_{\text{rmin}}$), $\sigma_{\text{tot}} \propto d$, where $d$ is the separation of the original binary. From Figure 1, we can see that for a smaller binary separation, a lower fraction of encounters with a third star will lead to a clean exchange. As fly-bys and exchanges tend to harden a binary (by an average amount of $\sim 40\%$), one might imagine the following scenerio for the evolution of a binary fed into the core of a globular cluster: the initial binary is sufficiently wide that clean exchanges or fly-bys are more likely than mergers, however after a small number of such encounters, the binary has been hardened sufficiently that mergers are more likely. Once a merger occurs, the merged object will either be unbound to the third star, and thus the binary can be considered to have been broken up, or it will be bound to the third star, forming what we dubbed a *merged binary*. Such a binary typically has a wide separation, and thus represents a large target for subsequent encounters with other stars.

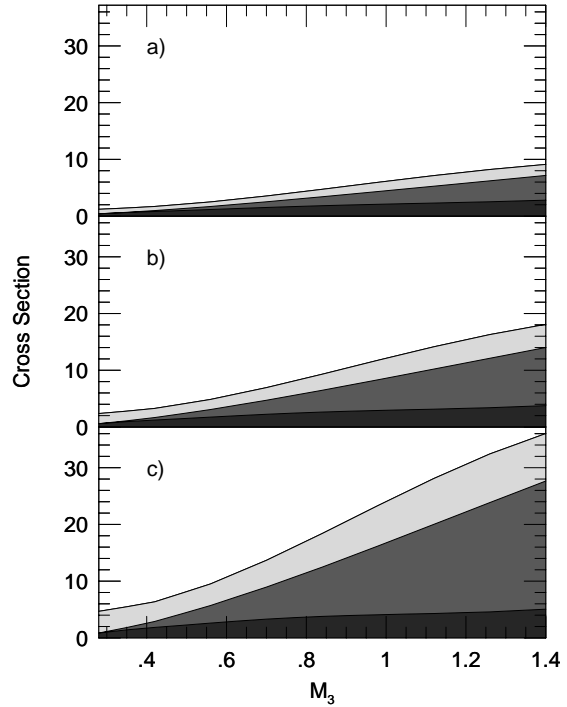

**Figure 1.** Relative cross sections for mergers, clean exchanges, and clean fly-bys (shaded black, dark grey, and light grey respectively) for encounters involving binaries containing two $0.7M_\odot$ stars of separations a) $50R_\odot$, b) $100R_\odot$, and c) $200R_\odot$. $M_3$ is given in solar units.

### 3.2 Encounter Rates for Particular Binaries

We next consider the encounter rate predicted from our computed cross sections. The rate of encounters between a given binary and the $i$th class of field stars, of number density, $n_i$, is given by $n_i \langle \sigma_{\text{bi}} V_{\text{bi}} \rangle$, where $V_{\text{bi}}$ is the relative speed at infinity between the binary and the field stars of the $i$th class, as given in equation (8). In order to compute the encounter rates as a function of field star mass, we must select a stellar population. In this general discussion, a simple power law will suffice, i.e. $dN/dM \propto M^{-\alpha}$. In Figure 2 we show the computed encounter rates for the binary of separation $50 R_\odot$ with values of $\alpha$ of 0, 1, and 2, with the rates normalised to the clean exchange cross section when $M_3 = 0.7 M_\odot$, and $\alpha = 0$. As in Figure 1, clean exchanges, mergers, and clean fly-bys are shaded black, dark grey, and light grey, respectively. The dependence of the relative encounter rates on $\alpha$ is clear; the flatter the stellar population, the greater the relative rates of clean exchanges compared to mergers. The IMF of the galaxy seems to have $\alpha \sim 2.35$, however the relatively heavy stars in a globular cluster preferentially sink to the core resulting in a flatter distribution, possibly even changing the sign of the distribution (Grabhorn et al. 1992).

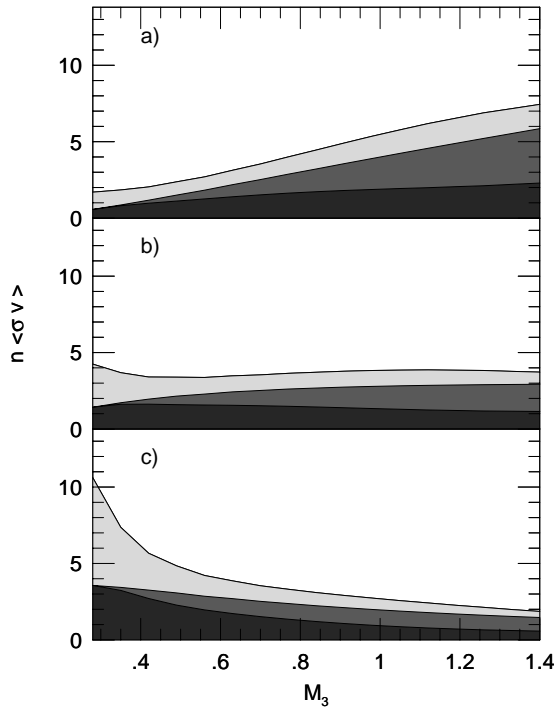

**Figure 2.** The relative rates for mergers, clean exchanges, and clean fly-bys (shaded black, dark grey, and light grey respectively) for encounters involving binaries containing two $0.7 M_\odot$ stars of separation $50 R_\odot$ for three stellar populations having values of the power law, $\alpha$ of a) 0, b) 1, and c) 2. $M_3$ is given in solar units.

## 4 PRODUCTS OF FIRST ENCOUNTERS

We now consider the products of first encounters in more detail. We consider encounters between the three different binaries and the three stellar populations considered in §3.2, leading to nine sets of runs in total. For each stellar population, we consider a total of $10^5$ stars and bin them logarithmically in mass. The number of stars contained in each bin, along with the average mass of the stars in that bin, is given in Table 5, for the three values of $\alpha$ used. We per-

**Table 5.** Number of stars contained in each mass bin for the three power laws considered.

| mass class | $M_{av}$ | Nature | $\alpha$ 0 | 1 | 2 |
|---|---|---|---|---|---|
| 1 | 0.1400 | MS | 2446.8 | 10000.0 | 24468.1 |
| 2 | 0.1808 | MS | 3160.2 | 10000.0 | 18944.8 |
| 3 | 0.2335 | MS | 4081.5 | 10000.0 | 14668.2 |
| 4 | 0.3016 | MS | 5271.5 | 10000.0 | 11357.1 |
| 5 | 0.3896 | MS | 6808.4 | 10000.0 | 8793.4 |
| 6 | 0.5031 | MS | 8793.4 | 10000.0 | 6808.4 |
| 7 | 0.6498 | MS | 5678.5 | 5000.0 | 2635.7 |
| 8 | 0.8393 | MS | 7334.1 | 5000.0 | 2040.8 |
| 9 | 0.6498 | WD | 5678.5 | 5000.0 | 2635.7 |
| 10 | 0.8393 | WD | 7334.1 | 5000.0 | 2040.8 |
| 11 | 1.0840 | WD | 18944.8 | 10000.0 | 3160.2 |
| 12 | 1.4000 | NS | 24468.1 | 10000.0 | 2446.8 |

The average mass for each mass bin, $M_{av}$, is given in solar units.

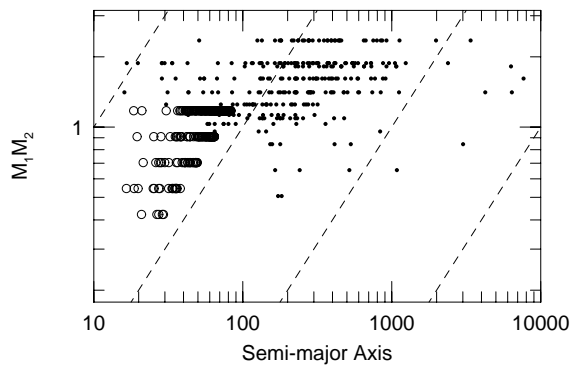

**Figure 3.** Clean-exchange (open circles) and merged binaries (black dots) produced in first encounters between a binary of separation, $d = 50 R_\odot$, and a population having a power-law distribution with $\alpha = 0.0$ (see §4). $M_1 M_2$ is the product of the masses of the two components, given in $M_\odot$, and the semi-major axis of the binary is given in $R_\odot$.

formed a monte carlo run of 10000 encounters for each of the nine combinations, counting up the number of clean exchanges, etc that occurred. The number of the various outcomes are listed in Table 6. From this table we see that, for a given value of $\alpha$, the number of clean exchanges occurring increases with increasing binary separation, $d$, as expected. The majority of clean exchanges involved either a white dwarf or neutron star as the incoming third star, which is as expected; more-massive stars are more effective at removing one of the components of the original binary in an exchange. We see in all cases, the vast majority of scattering induced mergers (SIMs) involve the merger of two main-sequence stars. This is because the most common form of a SIM involves the two components of the original binary merging; here we have assumed the original binaries were comprised of two main-sequence stars. The number of merged binaries

**Table 6.** The frequency of the possible outcomes for first encounters between a binary and a single star, as a function of the IMF power law, $\alpha$, and the separation of the binary, $d$.

| $\alpha$ | $d$ | FBs | CXs | | | SIMs | | | safe MBs | | | vulnerable MBs | | | SMOs | | |
|---|---|---|---|---|---|---|---|---|---|---|---|---|---|---|---|---|---|
| | | | MS | WD | NS | MS | WD | NS | MS | WD | NS | MS | WD | NS | MS | WD | NS |
| 0 | 50  | 2945 | 321 | 1204 | 1994 | 398 | 118 | 90 | 1456 | 714 | 620 | 102 | 19 | 19 | 12 | 10 | 3 |
| 0 | 100 | 3428 | 435 | 1497 | 2271 | 299 | 67  | 70 | 1005 | 490 | 413 | 143 | 41 | 21 | 6  | 3  | 0 |
| 0 | 200 | 3448 | 511 | 1696 | 2528 | 205 | 75  | 39 | 532  | 303 | 238 | 251 | 92 | 82 | 1  | 0  | 0 |
| 1 | 50  | 3400 | 411 | 1139 | 1320 | 442 | 126 | 57 | 1850 | 678 | 401 | 143 | 21 | 12 | 30 | 11 | 4 |
| 1 | 100 | 3818 | 557 | 1395 | 1517 | 320 | 78  | 40 | 1248 | 457 | 277 | 227 | 47 | 19 | 13 | 6  | 2 |
| 1 | 200 | 4106 | 580 | 1601 | 1640 | 238 | 59  | 27 | 739  | 264 | 161 | 348 | 90 | 47 | 2  | 0  | 0 |
| 2 | 50  | 4166 | 450 | 825  | 694  | 425 | 99  | 26 | 2342 | 560 | 199 | 192 | 15 | 7  | 33 | 13 | 1 |
| 2 | 100 | 4688 | 599 | 1029 | 808  | 343 | 66  | 13 | 1607 | 391 | 111 | 300 | 42 | 3  | 16 | 4  | 0 |
| 2 | 200 | 5079 | 727 | 1177 | 908  | 239 | 42  | 12 | 941  | 263 | 82  | 446 | 68 | 16 | 2  | 0  | 0 |

The initial binary was composed of two main-sequence stars of mass $0.8393M_\odot$. CXs are clean exchanges, SIMs are scattering-induced mergers (two of the three stars merged remaining unbound to the third star), MBs are merged binaries (two of the stars merged but remain bound to the third object), and SMOs are single merged objects (all three stars merge). Merged binaries are split into safe (those resilient to being broken up by a subsequent encounter) and vulnerable (those likely to be broken up on a subsequent encounter). Each category is further subdivided by the nature of the third star the binary interacted with: MS for main-sequence stars, WD for white dwarfs, and NS for neutron stars.

produced decreases with $d$. We also note that the fraction of merged binaries vulnerable to being broken up by a subsequent encounter with a third star increases with increasing $d$. This is no surprise; it is merely that merged binaries formed from wider binaries will themselves be wider. The number of single merged objects (SMOs) is small in all cases, with more being produced for smaller values of $d$. Most involve three main-sequence stars, encounters between the binary and massive compact objects (i.e. white dwarfs and neutron stars) tend to lead to either exchanges, or merged binaries. Increasing the value of the IMF power law, $\alpha$, leads to fewer neutron stars and white dwarfs, and thus the number of encounters involving such objects decreases.

In Figure 3, we plot $M_a M_b$ (where $M_a$ and $M_b$ are the masses of the components of the new binary) versus the semi-major axes for the clean-exchange, and merged binaries produced from encounters involving a binary of separation, $d = 50R_\odot$, and $\alpha = 0$. Lines of constant binding energy (i.e. $M_a M_b / d = $ a constant) are also included. In the original binary, $M_1 M_2 \sim 0.7$, hence we can clearly see how all the clean-exchange binaries are harder than the original binary, with most being only slightly harder. However, a large number of the merged binaries are less bound. Other binary separations and values of $\alpha$ produce figures similar to Figure 3. Increasing $\alpha$ (i.e. including fewer massive stars in the population) leads, unsurprisingly, to distributions in both populations of binaries more weighted to lower values of $M_a M_b$. Larger values of $d$ simply shifts both distributions towards greater values of semi-major axes.

In Figure 4, we plot the semi-major axis vs. eccentricity of the merged binaries and clean exchanges. Lines of constant pericentre separation are also included. From this figure, it is clear that a large fraction of post-encounter binaries will have appreciable eccentricity. We will discuss the timescale for circularisation in a later section. It is important to realise that the final separation $d$, after circularisation, can be considerably smaller than the initial semi-major axis of the eccentric binary, and is given by $d = (1-e^2)d_{\rm ecc}$. Hence if the initial eccentricity, $e \sim 0.9$, $d/d_{\rm ecc} \sim 0.2$.

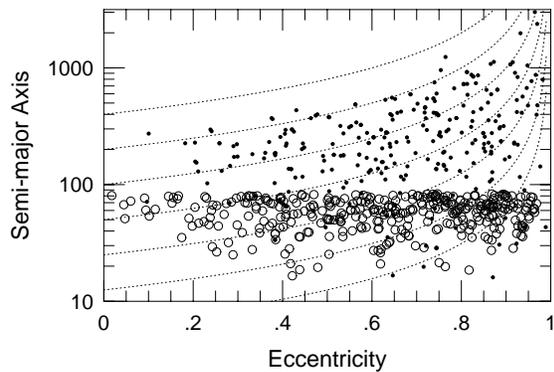

**Figure 4.** The semi-major axis (in solar units) and eccentricity of clean-exchange (open circles) and merged binaries (black dots) produced in first encounters between a binary of separation, $d = 50R_\odot$, and a population having a power-law distribution with $\alpha = 0.0$ (see §4).

## 5 MULTIPLE ENCOUNTERS

As discussed in §4, many of the systems produced in first encounters with primordial binaries will suffer subsequent encounters in a timescale shorter than a Hubble time, in some cases in a considerably shorter time. For sufficiently wide binaries, we have seen in §4 that most encounters will result in either a fly-by, or a clean exchange. In either case, the post-encounter binary will most often have been hardened (typically by $\sim 40\%$), the change in potential energy being seen in a boost to the binary's kinetic energy. The harder the binary, the greater the kick received, of course. After only a few such encounters, the kicks are large enough to remove the binary from the core. A subsequent encounter is thus delayed until the binary has diffused back into the

**Table 7.** Parameters for the globular cluster models considered.

| Model | $\alpha$ | $W_0$ | $r_0/\text{pc}$ | $r_c/r_0$ | $r_h/r_c$ | $r_t/r_c$ | $\rho_0$ | $\sigma_0$ | $\bar{m}_c$ | $M_{\text{tot}}$ | $M_c/M_t$ | $\tau_{\text{rc}}$ | $\delta$ |
|---|---|---|---|---|---|---|---|---|---|---|---|---|---|
| A | 2.00 | 12.0 | 0.73 | 1.22 | 11.14 | 63.40 | 2.E4 | 8 | 1.07 | 8.47E5 | 0.022 | 1.36E7 | 1.5 |
| B | 2.35 | 12.0 | 0.73 | 1.05 | 10.96 | 60.80 | 2.E4 | 8 | 0.93 | 7.70E5 | 0.025 | 1.65E7 | 1.5 |
| C | 2.70 | 12.0 | 0.73 | 0.93 | 11.16 | 59.68 | 2.E4 | 8 | 0.82 | 7.30E5 | 0.026 | 1.93E7 | 1.5 |
| D | 2.00 | 16.0 | 0.29 | 1.31 | 40.35 | 201.70 | 2.E5 | 10 | 1.24 | 2.24E6 | 0.005 | 1.93E6 | 2.2 |
| E | 2.35 | 16.0 | 0.29 | 1.20 | 32.70 | 163.54 | 2.E5 | 10 | 1.10 | 1.76E6 | 0.007 | 2.30E6 | 2.1 |
| F | 2.70 | 16.0 | 0.29 | 1.06 | 33.56 | 163.71 | 2.E5 | 10 | 0.96 | 1.65E6 | 0.007 | 2.66E6 | 2.1 |
| G | 2.35 | 8.0 | 1.74 | 0.88 | 3.97 | 20.95 | 2.E3 | 6 | 0.76 | 3.00E5 | 0.084 | 1.08E8 | 1.0 |
| H | 2.35 | 6.0 | 2.32 | 0.81 | 2.74 | 12.60 | 5.E2 | 4 | 0.67 | 1.00E5 | 0.147 | 1.65E8 | 0.9 |
| I | 2.35 | 4.0 | 2.75 | 0.72 | 2.05 | 7.79 | 2.E2 | 3 | 0.58 | 3.37E4 | 0.270 | 2.21E8 | 0.6 |

$\alpha$ is the exponent of the IMF power law, $W_0$ is the depth of the potential well (in unit of the central velocity dispersion $\sigma_0$). $r_0$ is the characteristic radius (in pc), $r_c$ is the core radius, $r_h$ is the half-mass radius, and $r_t$ is the tidal radius. The central density, $\rho_0$, is given in units of $M_\odot/\text{pc}^3$. The central velocity dispersion of stars of mass $\bar{m}_c$ (given in $M_\odot$), $\sigma_0$ is given in km/s. The total mass of the cluster, $M_t$ is given in $M_\odot$. $M_c$ is the mass of the cluster core. The core relaxation timescale, $\tau_{\text{rc}}$, is given in years. $\delta$ is the parameter fit for the gravitational potential (see equation [22]).

core again, this time being a very sensitive function of the apogee of the trajectory of the binary.

When discussing energetics of stars in globular clusters, it is convenient to work in units of $m\sigma^2 = kT$, where $m$ is the mean mass of stars in the core, and $\sigma$ is the characteristic 1-D velocity dispersion. For single-mass models of moderately concentrated clusters, the central potential $\sim 5 - 10kT$ (Binney & Tremaine 1987). For stars of equal mass, the third star carries off 2/3 of kinetic energy in the recoil, hence to eject the binary from the cluster we require $\Delta E \sim 30 - 60kT$ (as the binary contains two stars). Given that a binary typically hardens by about 40%, binaries are therefore vulnerable to ejection once they harden to $E_{\text{bin}} \sim 80 - 160kT$. The equivalent binary separation will depend on the velocity dispersion, and the masses of the components, but assuming $\sigma \sim 10$km/s, and stars of solar mass, yields a separation of $d \sim 20 - 40 R_\odot$.

## 5.1 Cluster Models

We use King-Mitchie models to describe various globular clusters. We model the stellar population by some power law, where the number of stars having a mass between $m$ and $m + dm$ is given by $dN(m) \propto m^{-\alpha}dm$. The stars are binned by mass into different classes $j$ and the stellar distributions are computed using a distribution function having a lowered Maxwellian energy dependence (King 1966), with an anisotropy term taking the form $\exp(-\beta J^2)$ (Mitchie 1962). For the $j$th mass bin, we have

$$f_j(E, J) \propto (e^{-A_j E} - 1)e^{-\beta J^2} \quad (9)$$

For details of the calculation of King-Mitchie models, see Gunn and Griffin (1979). The problem reduces to the solution of a first order system

$$\xi^2 \frac{dW}{d\xi} = -U, \quad \frac{1}{\xi^2}\frac{dU}{d\xi} = 9\tilde{\rho}(\xi, W) \quad (10)$$

where $\xi = r/r_0$, $W = \psi/\sigma^2$, $\tilde{\rho} = \rho/\rho_0$ and $U = 9\int_0^\xi \sigma \xi'^2 d\xi'$; $r_0$ being the scale radius, $\sigma$ is the velocity dispersion of stars of mass, $\bar{m}_c$ ($\bar{m}_c = \tilde{\rho}_0^{-1}\sum m_j \tilde{\rho}_j$). The central velocity dispersion is denoted $\sigma_0$ and $\rho_0$ is the central density. The normalised density at a given radius, $\tilde{\rho}(\xi, W) = \sum \tilde{\rho}_j(\xi)$, where the density contribution of each mass class of stars, $\tilde{\rho}_j(\xi) = \alpha_j I_1(\mu_j W, \xi/\xi_a)/I_1(\mu_j W_0, 0)$ where $\xi_a = r_a/r_0$, $r_a$ being the *anisotropy* radius, $\mu_j = m_j/\bar{m}$, and $I_1(x, y) = \int_{|u|<(2x)^{1/2}} \exp(-\frac{1}{2}y^2 u_\perp^2) \times [\exp(-\frac{1}{2}u^2 + x) - 1]d^3u$. $\alpha_j$ is thus the contribution to the central density from mass bin $j$. In the models we consider here, we simplify the models by assuming the velocity dispersion is isotropic for the whole of a cluster (i.e. $r_a = \infty$). In a companion work (Davies & Benz [1994]), we consider models for the globular clusters for $\omega$ Cen and 47 Tuc produced by Meylan (1987, 1989) which include a non-trivial anisotropy radius.

The value of the scale radius, $r_0$, is given by

$$r_0 = \sqrt{\frac{9\sigma_0^2}{4\pi G \rho_0}} \quad (11)$$

The total mass of the system, $M_{\text{tot}}$ is given by

$$M_{\text{tot}} = \left(\frac{r_0 \sigma_0^2}{G}\right) U_{\text{tot}} \quad (12)$$

where $U_{\text{tot}} = 9\int_0^{\xi_t} \sigma \xi'^2 d\xi'$.

With $W_0$, $\xi_a$, and $\alpha_j$ for all the mass bins specified, equation (10) can be integrated until $W = 0$ at which point we have reached the cluster edge. One adjusts the values of $\alpha_j$ until the fractions of mass contained in the various mass bins agrees with the assumed mass function. Thus we are able to compute the density of each mass bin as a function of radius given certain characteristics of the globular cluster.

A choice of $\sigma_0^2$ and $\rho_0$ then yield a physical value for $r_0$, and $M_{\text{tot}}$. It should be noted that $r_0$ *and the core radius,* $r_c$ *are not necessarily identical.* The core radius is defined *observationally* as the radius at which the surface brightness is half the central value. The scale radius takes its definition from equation (11) and is the core radius for stars of mass $\bar{m}_c$. The two radii will take different values as long as the turn-off mass (the mass of the stars contributing mostly to the surface brightness) differs from $\bar{m}_c$. The difference is computed here, though the discrepancy is relatively small, always being less than 30%.

Table 8. Number of stars contained in each mass bin for the cores of the nine globular cluster models.

| mass class | $M_{av}$ | Nature | A | B | C | D | Model E | F | G | H | I |
|---|---|---|---|---|---|---|---|---|---|---|---|
| 1 | 0.1400 | MS | 900 | 1872 | 3239 | 86 | 234 | 416 | 8628 | 7757 | 6736 |
| 2 | 0.1808 | MS | 787 | 1528 | 2489 | 78 | 198 | 334 | 6673 | 5826 | 4905 |
| 3 | 0.2335 | MS | 720 | 1327 | 2037 | 76 | 183 | 296 | 5327 | 4498 | 3666 |
| 4 | 0.3016 | MS | 711 | 1245 | 1844 | 81 | 189 | 303 | 4472 | 3542 | 2729 |
| 5 | 0.3896 | MS | 778 | 1330 | 1929 | 101 | 232 | 378 | 4013 | 2938 | 2104 |
| 6 | 0.5031 | MS | 992 | 1686 | 2438 | 155 | 363 | 621 | 3952 | 2572 | 1670 |
| 7 | 0.6498 | MS | 1549 | 2632 | 3732 | 320 | 776 | 1405 | 4356 | 2428 | 1370 |
| 8 | 0.8393 | MS | 3044 | 4873 | 6164 | 949 | 2286 | 4006 | 5278 | 2473 | 1185 |
| 9 | 0.6498 | WD | 1797 | 2420 | 2736 | 371 | 714 | 1030 | 4005 | 2232 | 1260 |
| 10 | 0.8393 | WD | 2098 | 2422 | 2222 | 654 | 1136 | 1445 | 2623 | 1229 | 589 |
| 11 | 1.0840 | WD | 4602 | 3934 | 2597 | 2610 | 3365 | 2964 | 2881 | 1167 | 478 |
| 12 | 1.4000 | NS | 4259 | 1929 | 724 | 4890 | 2699 | 1158 | 1110 | 414 | 153 |
| total | | | 22237 | 27198 | 32152 | 10372 | 12376 | 14357 | 53320 | 37074 | 26846 |

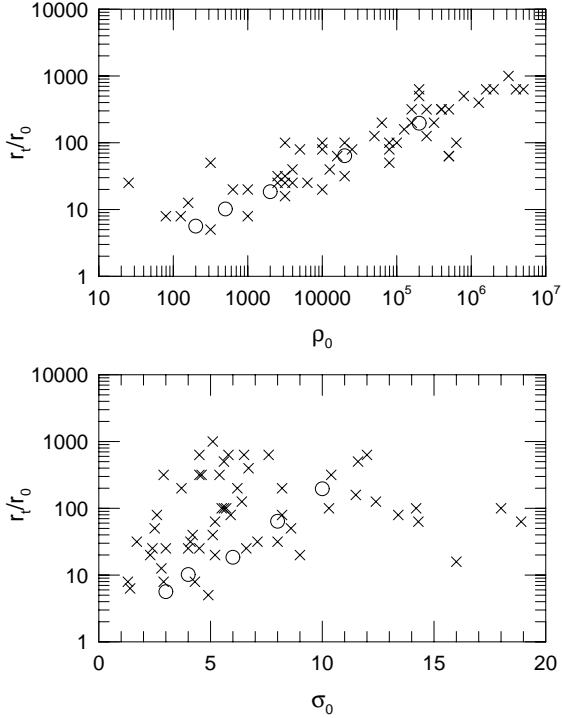

**Figure 5.** The tidal radius divided by the scale radius, $r_t/r_0$, as a function of a) $\rho_0$ and b) $\sigma_0$ for the 56 clusters modelled by Pryor & Meylan (1993), labelled with "×"s and for our cluster models with $\alpha = 2.35$ (labelled with open circles).

We considered three power laws for the IMF, taking values of $\alpha$ of 2.00, 2.35, and 2.70. We considered stars having a mass between $0.14 M_\odot$ and $15.0 M_\odot$. Stars more massive than $8.0 M_\odot$ were considered to evolve into neutron stars via type II SN. Those between 4.7 and $8.0 M_\odot$ are deemed to leave no remnant (Iben & Renzini 1983), whilst the remaining stars above the present turn-off leave white dwarfs of mass, $M_{wd} = 0.58 + 0.22(M_{ms} - 1) M_\odot$. The stars are then binned in the same manner as the populations listed in Table 5.

We considered values of the central potential depth, $W_0$, ranging from 4 to 16. For a given IMF, we solved equation (10) to produce the spatial distribution of the stars in each mass class. Suitable values for $\sigma$ and $\rho_0$ were then chosen by comparing our model clusters with properties derived for 56 clusters by Pryor & Meylan (1993). In Figure 5 we plot the concentration, $r_t/r_0$, as a function of $\rho_0$ and $\sigma$ for the 56 clusters studied by Pryor & Meylan. The values adopted for our models for $\alpha = 2.35$ are also shown ($r_t/r_0$ was found to be largely independent of $\alpha$ for the range of IMFs we considered). Having selected values for $\rho_0$ and $\sigma$, the values for $r_0$ and $M_{tot}$ were computed, and are listed in Table 7, together with other properties of the globular cluster models.

To obtain the total stellar population in the core, we integrated the number densities of the various stellar species out to a scale radius. The stellar population in the core of a cluster differs from the IMF of the whole cluster; mass segregation results in the core containing more massive stars. This effect is more pronounced in more collapsed clusters (i.e. clusters with a larger value of $W_0$), as can be seen in Table 8.

### 5.2 Encounters between Two Single Stars

Before we consider encounters between binaries and single stars, we consider the rate of encounters between two single stars, given the core models in Table 8. The rate of encounters between two stellar species, 1 and 2, passing within some minimum distance, $R_{min}$ is given by

$$\Gamma_{12} = 6 \times 10^{-9} \frac{n_1}{10^4 \text{ pc}^{-3}} \frac{n_2}{10^4 \text{ pc}^{-3}} \frac{M_1 + M_2}{M_\odot}$$
$$\times \frac{R_{min}}{R_\odot} \frac{10 \text{kms}^{-1}}{V_\infty} \text{ yr}^{-1} \text{pc}^{-3} \qquad (13)$$

where $n_1$ and $n_2$ are the number densities of the two species. We use an analytic expression for $R_{min}$ suggested by Benz and Hills (1991):

$$\frac{R_{min}}{R_1} = 3.28 + 0.45 \log\left(\frac{M_2}{M_1}\right) + 0.22 \log\left(\frac{M_1}{M_\odot} \cdot \frac{R_\odot}{R_1}\right) \qquad (14)$$

**Table 9.** The production rates of blue stragglers (BS), white dwarf/main-sequence star (WD-MS), and neutron star/main-sequence star (NS-MS) systems via encounters between single stars in various globular cluster models.

| Model | $W_0$ | $\alpha$ | BS | WD-MS | NS-MS |
|-------|-------|----------|-------|-------|-------|
| A | 12.0 | 2.00 | 0.372 | 0.931 | 0.666 |
| B | 12.0 | 2.35 | 1.124 | 1.656 | 0.537 |
| C | 12.0 | 2.70 | 2.264 | 1.970 | 0.291 |
| D | 16.0 | 2.00 | 0.242 | 1.250 | 2.260 |
| E | 16.0 | 2.35 | 1.489 | 4.489 | 3.186 |
| F | 16.0 | 2.70 | 4.876 | 8.521 | 2.566 |
| G | 8.0  | 2.35 | 0.230 | 0.272 | 0.096 |
| H | 6.0  | 2.35 | 0.123 | 0.062 | 0.009 |
| I | 4.0  | 2.35 | 0.047 | 0.017 | 0.002 |

Encounter rates are given in units of encents per $10^8$. $W_0$ is the depth of the cluster potential well (in units of the central velocity dispersion), and $\alpha$ is the exponent of the IMF power law.

We thus calculate the encounter rates between the various stellar species for the models listed in Table 8; the results being listed in Table 9, with production rates being given in units of encounters/$10^8$ years. We assume here that all close encounters between two main-sequence stars that become bound produce blue stragglers. There has been much debate concerning whether binaries can be formed via tidal capture. Here we simply list all close encounters between main-sequence stars and white dwarfs and neutron stars. Some fraction of these encounters may lead to binaries rather than a smothered compact object. Simulations suggest that less than half the encounters produce binaries (Davies, Benz & Hills 1992). From Table 8, we see that the encounter rates are an extremely strong function of cluster concentration. The most-collapsed clusters (models D, E, and F) show larger rates. The rates are also seen to be a function of the IMF; smaller values of the power-law exponent, $\alpha$, implies a larger fraction of neutron stars and white dwarfs compared to main-sequence stars, and this difference is echoed in the encounter rates.

### 5.3 Timescales

The components of a tight binary will spiral-in because of angular momentum losses via magnetic winds and gravitational radiation. For the binaries containing at least one main-sequence, it has been suggested that magnetic breaking will dominate over general relativistic effects (Verbunt & Zwaan 1981). The timescale for two stars to spiral-in from a distance $d_{\rm init}$ to a separation $d_{\rm final}$ is given by (DiStefano & Rappaport 1992)

$$\tau_{\rm mb} = 4.3 \times 10^5 \left(\frac{M_2}{M_\odot}\right)\left(\frac{R_\odot}{R_1}\right)^\gamma \left(\frac{M_\odot}{M_1+M_2}\right)^2 \times (d_{\rm init}^4 - d_{\rm final}^4) \text{ years} \quad (15)$$

where here we will take $\gamma = 1$.

For systems containing two compact objects, inspiral will still occur due to angular momentum loss from gravitational radiation. Beginning from a separation $d$, merger will occur on a timescale given by (Landau & Lifshitz 1962)

$$\tau_{\rm gr} = 1.55 \times 10^8 \left(\frac{M_1}{M_\odot}\right)^{-1}\left(\frac{M_2}{M_\odot}\right)^{-1}\left(\frac{M_1+M_2}{M_\odot}\right)^{-1} \times \left(\frac{d}{R_\odot}\right)^4 \text{ years} \quad (16)$$

For binaries, of separation $d$, containing a convective star of mass $M_1$ and radius $R_1$, and a compact object of mass $M_2$, the timescale for the circularisation of the binary is given by (DiStefano & Rappaport 1992)

$$\tau_{\rm circ} = 2.7 \left(\frac{M_1}{M_2}\right)\left(\frac{M_1}{M_1+M_2}\right)\left(\frac{d}{R_1}\right)^8 \text{ years} \quad (17)$$

If a binary circularizes, the final separation, $d = (1-e^2)d_{ecc}$ where $d_{ecc}$ is the semi-major axis of the eccentric binary, having eccentricity, $e$.

The timescale for a binary of separation $d$ to encounter something is largely independent of the slope of the IMF (see Figure 2). It can be approximated by

$$\tau_{\rm enc} \sim 3 \times 10^{11} \frac{1}{n_4} \left(\frac{d}{R_\odot}\right)^{-1} \text{ years} \quad (18)$$

where $n_4$ is the total number density of stars in units of $10^4/\text{pc}^3$.

For computational simplicity, the gravitational potential for one of the globular cluster models can be approximated (for $r_c < r < 10^\delta r_c$) as

$$\Phi \sim \frac{\Phi_0}{\delta} \log\left(\frac{r}{r_c}\right) - \Phi_0 \quad (19)$$

where $\Phi_0$ is the depth of the potential. $\delta$ was determined empirically for each model, and is listed in Table 7.

The relaxation time can be taken to be $\tau_r = \tau_{\rm rc}(r/r_c)^2$ where the core relaxation time, $\tau_{\rm rc}$ is given by (Lightman & Shapiro 1977)

$$\tau_{\rm rc} = 1.55 \times 10^7 \frac{(M_\odot/m)}{\log(0.5M_t/m)}\left(\frac{\sigma}{km/s}\right)\left(\frac{r_c}{pc}\right)^2 \quad (20)$$

where $m$ is the mean stellar mass in the core, and $M_t$ is the total mass of the cluster, as listed in Table 7.

Using the assumptions discussed above for the potential well of a globular cluster and the radial dependence of the local relaxation time, we can compute the timescale for the return of a binary to the core of a cluster, having been removed with specific energy $\epsilon_{\rm kick}$. Following Hut, McMillan & Romani (1992), we assume the evolution of the orbital apocenter, $r_a$, is given by

$$\frac{1}{r_a}\frac{dr_a}{dt} = \frac{1}{\tau_r(r_{\rm eff})} \quad (21)$$

where $r_{\rm eff} = (r_c^3 r_a)^{1/4}$, and $\tau_r$ is the local relaxation time. Intergration and simple algebra then gives the timescale for return to the core as

$$\tau_{\rm ret} \sim \frac{2}{3}\tau_{\rm rc}\left(10^{(3\delta\epsilon_{\rm kick}/2\Phi_0)} - 1\right) \quad (22)$$

These various timescales are plotted in Figure 6, for three different cluster models. For simplicity here we assumed all stars were solar. In all three plots, the timescales for inspiral via magnetic winds and gravitational radiation are unchanged, as is the timescale for binary circularisation, as these depend purely on the properties of the binary, of course. Both the encounter timescale and the timescale to return to the core, assuming the binary received a kick in its last encounter, increase with decreasing density. For a binary of separation, $d = 1000R_\odot$, the encounter timescale ranges from $\sim 10^7$ years for the densest cluster shown in Figure 6, to $\sim 10^9$ years for the least dense. As we shall see later, this range in timescales will turn out to be crucial in determining what fraction of binaries will have suffered many encounters in a cluster core, leading often to ejection, merging, or breakup. Imagine that we begin with a binary of separation, $d = 1000R_\odot$. From Figure 6, we note that in the densest two cluster cores, the binary will quickly encounter a third star. In such wide binaries, most encounters will either be a fly-by or a clean exchange. In either outcome, we will be left with a detached, but hardened, binary. Typically, we expect the binary to be hardened by 40%, in other words we expect $d_{\text{new}} \simeq 0.7 d_{\text{old}}$. As for the two densest clusters, $\tau_{\text{enc}} \ll 10^{10}$ years, we expect this binary to undergo many subsequent encounters, gradually getting harder. As the binary gets harder, encounters with a third star are more likely to lead to a merger. We have seen that merged binaries tend to have much larger separations, hence we expect any merged binaries to rapidly undergo a subsequent encounter (on a timescale $\sim 10^7 - 10^8$ years for the two densest clusters considered in Figure 6). We note from Figure 6a that $\tau_{\text{enc}} \sim 10^{10}$ years when $d \sim 100R_\odot$. We are therefore unlikely to produce many binaries much harder than this in such low-density clusters, and in turn we expect the number of merged systems to be reduced.

It is also important to note that $\tau_{\text{ret}} \ll \tau_{\text{enc}}$ for $d \geq 10 - 20R_\odot$. $\tau_{\text{ret}}$ increases rapidly with decreasing $d$ for $d \sim 10 - 20R_\odot$, as binaries of this separation will get ejected from the core almost to the outer edge of the cluster halo, and thus take an extremely long time to diffuse back into the core. It is also worth noting that in the three clusters considered in Figure 6, subsequent encounters on timescales $\leq 10^{10}$ years are not delayed significantly by having to wait for the binary to re-enter the core.

We see from Figure 6 that magnetic breaking will bring a $1M_\odot$ main-sequence star and its $1M_\odot$ companion together in $\leq 10^{10}$ years, if $d \leq 10R_\odot$. Gravitational radiation performs a similar function when the initial separation $d \leq 4R_\odot$. In order to reach such small initial separations, we will first have to pass through a common envelope phase, as will be discussed in the next section.

We note also that $\tau_{\text{circ}} \leq 10^{10}$ years for $d \leq 20R_\odot$. Recalling that after circularisation, $d = (1 - e^2)d_{\text{ecc}}$, we see that relatively wide, eccentric binaries, ultimately evolve into much smaller, circular, ones. For example, a binary with $e = 0.8$, and $d_{\text{ecc}} = 30R_\odot$ will produce a circular binary, with $d \simeq 10R_\odot$. In other words, if we can produce fairly hard binaries ($d \sim 10 - 40R_\odot$) with sizeable eccentricity, these may then circularise into systems hard enough to come into contact in a Hubble time.

### 5.4 Stellar Evolution Issues

A white dwarf smothered by the remnant of a main-sequence star will evolve into a red giant on a thermal timescale ($10^6 - 10^7$ years). Such objects contained in resilient merged binaries will engulf the other component of the binary form-ing a common envelope system. In such a system, the white dwarf and other component will spiral together as the common gaseous envelope is ejected. The final separation of the two stars may be estimated by simply equating the binding energy of the envelope with the change in binding energy of the two stars, within some efficiency $\alpha_{ce}$. We thus have

$$\begin{aligned} E_{\text{env}} &= \alpha_{ce} \Delta E_g \\ \text{where } E_{\text{env}} &= \frac{G(M_c + M_e)M_e}{\lambda a_i r_l} \\ \text{and } \Delta E_g &= \frac{GM_2 M_c}{2a_f} - \frac{GM_2(M_e + M_c)}{2a_i} \end{aligned} \quad (23)$$

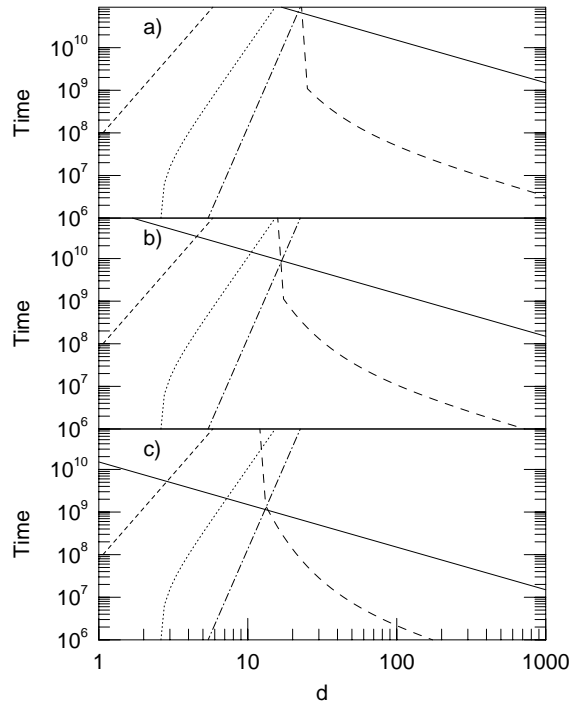

**Figure 6.** The various timescales (in years) for the cores of three globular cluster models as a function of the binary separation, $d$ (given in solar units). The three models have different central densities: a) $2000M_\odot/\text{pc}^3$, b) $2 \times 10^4 M_\odot/\text{pc}^3$, and c) $2 \times 10^5 M_\odot/\text{pc}^3$. The timescale for an encounter with a single star is shown with a solid line, the timescale to return to the core after being ejected by an encounter is given with a long-dashed line. The timescale for circularisation is given by a dot-dashed line, whilst the timescales for inspiral due to magnetic breaking and gravitational radiation are given by dotted and short-dashed lines respectively.

Where $M_e$ and $M_c$ are the envelope and core masses, $M_2$ is the mass of the second star, $a_i$ and $a_f$ are the initial and final separations, $r_l$ is the Roche lobe radius of the red giant, and $\lambda$ depends on the structure of the red giant envelope and $\sim 0.5$. Hence with a choice of $\alpha_{ce}$ (typically taken to be 0.2-0.4 [Taam & Bodenheimer 1989]), we are able to compute the final separation of the white dwarf and second star. Such a mechanism will be of great use in producing hard binaries in globular clusters.

Next we consider neutron stars smothered by main-sequence stars. The subsequent evolution of such an object is extremely uncertain. It is possible that a stable envelope configuration can be obtained where material is accreted onto the surface of the neutron star and supports the envelope. Such a *Thorne-Zytkow Object* will live for $\sim 10^6$ years as a supergiant before the envelope has been completely ejected via a wind (Cannon 1993). If we assume this evolutionary scenario is correct, then any smothered neutron stars contained in binaries will lead to common envelopes as the envelope expands. The evolution can then be modelled as described above for smothered white dwarfs. It has also been suggested that the envelope will be ejected after an extremely small amount of material has been accreted. If this latter scenerio is correct, then no common envelope phase will occur, and the merged binaries are likely to broken up by the instantaneous mass loss, or the binaries will be so weakly bound after the episode of mass loss that they are likely to be broken up by an encounter with a passing third star.

### 5.5 The Evolution of Binaries in Globular Cluster Cores

Following Hut, McMillan & Romani (1992), we consider the core of a globular cluster to be composed of a static population of single stars and a population of evolving binaries. In practice, the background of single stars will also be evolving, though the core may be supported from collapse by the energy released from the binaries. This analysis represents a reasonable intermediate step until a full n-body treatment is conducted. We imagine binaries diffusing into the core, encountering a single star (or perhaps many single stars) and producing some new binary/object. We neglect 2+1 encounters occuring outside of the core (this is not unreasonable as only a small fraction of the encounters will occur outside of the core). For a given binary, we compute its cross sections for interactions with all the different mass classes of single stars, select one at random (weighted by the cross sections) thus producing a new object. If the new system is a binary that is not ejected from the cluster, we repeat the process, thus simulating subsequent encounters. If the new binary is removed from the core, but not from the cluster, we estimate a timescale for the binary to return to the core, as given by equation (22), and reinsert it into the core after the computed delay.

We injected 1000 binaries into the core of each of the cluster models and evolved the binaries until one of the following occurred: 1) a merged object was produced in a scattering-induced merger (SIM), 2) the binary was deemed to be likely to be broken up through an encounter with a third star ($M_1 M_2/d < 0.001$ in solar units), 3) a single merged object (SMO) was produced, 4) the binary was ejected from the cluster, or 5) the binary was deemed to have come into contact via inspiral from either a magnetized wind or gravitational radiation. Additionally, in some runs we halted the evolution of the binaries when a smothered neutron star was produced, as the subsequent evolution of such an object is currently unclear.

There are clearly a great number of parameters we can vary; the cluster model, described by its IMF, and concentration, the population of binaries we feed into the cluster core, and their distribution of injection times. We also have to decide how long to evolve the cluster binaries for. Finally, we have to consider two stellar evolution issues; assigning a value to the efficiency of the common envelope phase, $\alpha_{ce}$, and deciding how to treat smothered neutron stars (see §5.4). Rather than consider the full range of possibilities here, we show a restricted set, changing only one of the above parameters at a time in an attempt to show trends.

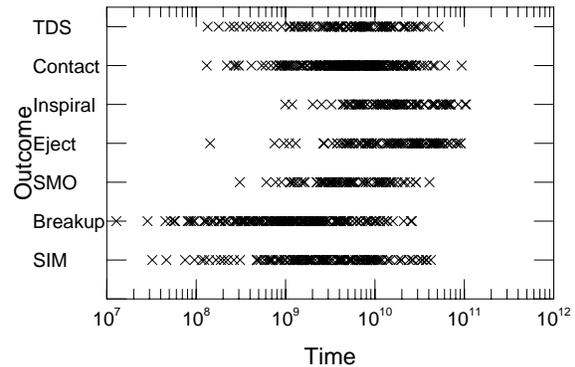

**Figure 7.** Final outcome of the binaries evolved in Run 1, as a function of when the final object was produced. SIM are scattering-induced mergers, SMOs are single merged objects, TDS refers to smothered neutron stars (thick disk systmes). Contact refers to binaries brought together via magnetic breaking, whilst inspiral refers to binaries brought together via gravitational radiation.

### 5.6 The Effect of Changing the Injection Time and Duration of the Run

We will begin by considering a moderately concentrated cluster ($W_0 = 12.0$) with an IMF, $\alpha = 2.35$. In run 1, we injected 1000 binaries (each containing two $0.8393 M_\odot$ main-sequence stars) into the core and let them evolve until one of the criteria described above were met. The initial binary separations were chosen randomly with equal probability in $\log d$ from the range $100 R_\odot < d < 704 R_\odot$ (the upper being by forcing $M_1 M_2/d > 0.001$ in solar units; this being the criteria for a hard binary unlikely to be broken up by a passing third star). In this run, we stop the evolution of a binary as soon as a smothered neutron star is produced, labelling the outcome as a TDS. We also adopted $\alpha_{ce} = 0.4$. In Figure 7 we plot the final outcomes of the 1000 binaries as a function of when the outcomes occurred. The first thing to note from this plot is that a fraction of the binaries are not resolved until much later than 15Gyr after they have been inserted into the cluster core. We see from Figure 7 that breakup and SIM events occur almost immediately with a large fraction of such events occurring after a few Gyr. However, we see that ejection and binary coalescence events (labelled as contact and inspiral in the figure) tend to occur later. This is not surpising; in the case $W_0$ is the depth of the cluster potential (in units of the central velocity dispersion). $\alpha$ is the exponent of the power-law IMF. $\alpha_{ce}$ is the efficiency in the common envelope stage. BP is the binary population: in type A, the initial binaries always contain two $0.8393 M_\odot$ main-sequence stars, in type B they contain two main-sequence stars drawn independently from the IMF, with the extra criterian that their total mass $> 1 M_\odot$. $t_{evol}$ is the total time the binaries were allowed to evolve (in Gyr), with the binaries being initially injected into the core at a time drawn randomly between zero and $t_{inj}$, given in (Gyr). BS is the number of blue stragglers produced when two main-sequence stars collided during an encounter.

**Table 10.** The outcomes of the evolution of 1000 binaries in various cluster models.

| Run | $W_0$ | $\alpha$ | $\alpha_{ce}$ | BP | $t_{evol}$ | $t_{inj}$ | BS | TDS | contact binaries | | | | | | MGRI | | |
|---|---|---|---|---|---|---|---|---|---|---|---|---|---|---|---|---|---|
| | | | | | | | | | XB | CV | SN | SW | BS | NW | $WD_a^2$ | $WD_b^2$ | $NS^2$ |
| 1 | 12.0 | 2.35 | 0.4 | A | $\infty$ | 0.0 | 343 | 157 | 44 | 77 | 5 | 54 | 93 | 33 | 51 | 1 | – |
| 2 | 12.0 | 2.35 | 0.4 | A | 15.0 | 0.0 | 365 | 121 | 16 | 69 | 2 | 53 | 76 | 21 | 30 | 0 | – |
| 3 | 12.0 | 2.35 | 0.4 | A | 15.0 | 15.0 | 295 | 86 | 16 | 32 | 4 | 42 | 34 | 5 | 11 | 0 | – |
| 4 | 12.0 | 2.35 | 0.4 | B | 15.0 | 15.0 | 191 | 66 | 9 | 32 | 1 | 17 | 48 | 11 | 8 | 0 | – |
| 5 | 12.0 | 2.70 | 0.4 | B | 15.0 | 15.0 | 305 | 34 | 10 | 38 | 1 | 32 | 50 | 2 | 6 | 0 | – |
| 6 | 12.0 | 2.00 | 0.4 | B | 15.0 | 15.0 | 107 | 105 | 22 | 20 | 0 | 15 | 25 | 5 | 11 | 0 | – |
| 7 | 12.0 | 2.35 | 0.4 | B | 15.0 | 15.0 | 188 | – | 22 | 33 | 0 | 14 | 52 | 10 | 11 | 0 | 1 |
| 8 | 12.0 | 2.35 | 0.4 | A | 15.0 | 15.0 | 273 | – | 22 | 30 | 5 | 42 | 56 | 14 | 15 | 0 | 0 |
| 9 | 12.0 | 2.35 | 0.4 | A | $\infty$ | 15.0 | 361 | – | 63 | 86 | 12 | 75 | 73 | 77 | 54 | 1 | 16 |
| 10 | 12.0 | 2.35 | 0.4 | B | $\infty$ | 15.0 | 272 | – | 52 | 85 | 4 | 58 | 79 | 74 | 57 | 0 | 18 |
| 11 | 12.0 | 2.35 | 0.2 | B | 15.0 | 15.0 | 186 | – | 9 | 31 | 0 | 21 | 44 | 17 | 10 | 0 | 5 |
| 12 | 12.0 | 2.35 | 0.1 | B | 15.0 | 15.0 | 185 | – | 16 | 24 | 1 | 17 | 42 | 18 | 32 | 0 | 4 |
| 13 | 12.0 | 2.35 | 0.05 | B | 15.0 | 15.0 | 178 | – | 17 | 24 | 2 | 11 | 39 | 31 | 35 | 0 | 7 |
| 14 | 16.0 | 2.35 | 0.05 | B | 15.0 | 15.0 | 168 | – | 24 | 31 | 2 | 9 | 19 | 138 | 42 | 0 | 66 |
| 15 | 16.0 | 2.35 | 0.4 | B | 15.0 | 15.0 | 144 | – | 47 | 37 | 0 | 7 | 23 | 86 | 31 | 0 | 28 |
| 16 | 16.0 | 2.35 | 0.4 | B | 15.0 | 15.0 | 169 | 220 | 22 | 29 | 0 | 9 | 29 | 34 | 29 | 0 | – |
| 17 | 16.0 | 2.35 | 0.4 | B | 15.0 | 0.0 | 150 | 287 | 21 | 34 | 0 | 9 | 30 | 39 | 28 | 0 | – |
| 18 | 16.0 | 2.00 | 0.4 | B | 15.0 | 15.0 | 86 | 223 | 27 | 12 | 1 | 2 | 8 | 19 | 9 | 0 | – |
| 19 | 16.0 | 2.70 | 0.4 | B | 15.0 | 15.0 | 280 | 160 | 13 | 49 | 4 | 33 | 54 | 35 | 55 | 0 | – |
| 20 | 16.0 | 2.00 | 0.4 | B | 15.0 | 15.0 | 83 | – | 39 | 13 | 0 | 1 | 16 | 40 | 3 | 0 | 41 |
| 21 | 8.0 | 2.35 | 0.4 | B | 15.0 | 15.0 | 65 | 2 | 1 | 2 | 1 | 7 | 19 | 0 | 0 | 0 | 0 |
| 22 | 6.0 | 2.35 | 0.4 | B | 15.0 | 15.0 | 31 | 0 | 0 | 0 | 0 | 0 | 10 | 0 | 0 | 0 | 0 |
| 23 | 4.0 | 2.35 | 0.4 | B | 15.0 | 15.0 | 13 | 0 | 0 | 1 | 0 | 0 | 2 | 0 | 0 | 0 | 0 |

TDS is the number of smothered neutron stars produced (when this was one of the criteria for terminating the evolution of the binary). Contact binaries are those systems brought into contact via angular momentum loss by magnetic winds. This outcome is further subdivided into: X-ray binaries (XB) when the system contains a neutron star more massive then its main-sequence companion, cataclysmic variables (CV) when the system contains a white dwarf more massive than its main-sequence companion, smothered neutron stars (SN) and smothered white dwarfs (SW) when the compact objects are less massive than their companions, and blue stragglers (BS) when both components are main-sequence stars. Systems brought into contact via gravitational radiation are labelled MGRI (mergers through gravitational radiation inspiral). This outcome is further subdivided into: neutron star/white dwarf mergers (NW), white dwarf/white dwarf mergers ($WD^2$), and neutron star/neutron star mergers ($NS^2$). $WD^2$ are further subdivided into those where the total mass exceeds the Chandrasekhar mass ($WD_a^2$) and those of smaller total mass ($WD_b^2$).

of ejections, the binary must first harden via exchanges and fly-bys until the recoil is sufficient to remove it from the cluster. To produce a coalescence via inspiral from gravitational radiation (labelled simply as inspiral in the figure), the binary first passes through a CE phase, the two compact objects being left $\sim 3 R_\odot$ apart. These two stars then spiral together as angular momentum is lost through gravitational radiation on a timescale $\sim 10$ Gyr. Binaries may also be brought into contact as angular momentum is lost via a magnetized wind from a main-sequence star component of the binary, on a timescale $\sim 1 - 5$ Gyr, if the initial separation $\sim 8 R_\odot$. Single-merged objects (SMOs) are produced only for encounters involving hardened binaries hence are only seen after $\sim 1 - 10$ Gyr.

In Table 10 we list the frequencies of the various possible outcomes. For the systems brought into contact by angular momentum loss from a magnetized wind we differentiate between five possible outcomes: if a main-sequence star is filling its roche lobe and transferring material to a neutron star an LMXB will be produced if the neutron star is the more massive of the two stars, otherwise a thick-disk system (TDS) will be produced with the neutron star smothered by the remains of the main-sequence star; we apply a similiar criterion for mass transfer onto a white dwarf, producing either a CV if the white dwarf is more massive, or a smothered white dwarf. This approximation is reasonable given the nature of these simulations. Finally, if both components of the binary are main-sequence stars, we tag the merged object as a blue straggler. This is another mechanism for producing blue stragglers in addition to those produced di-rectly through collisions of two main-sequence stars during a 2+1 encounter.

For systems coalescing after losing angular momentum via gravitational radiation, we distinguish between the following outcomes: the system may contain two neutron stars, a neutron star and a white dwarf, or two white dwarfs. In the latter case, we further distinguish between those systems containing a total mass greater than the Chandrasekhar limit ($1.4M_\odot$) and those less massive.

Run 2 is identical to run 1 except that we only evolved the binaries for 15 Gyr. Of the 1000 binaries, 212 of them were unresolved at the end of the run (in other words, they had not reached some end state, such as merging or ejection). In run 3, we injected the binaries at some random time drawn evenly between zero and 15 Gyr, and evolved them up to a time of 15 Gyr. This procedure increased the number of unresolved systems to 474. In other words, in a fairly concentrated cluster, a large fraction of the binaries that have entered the core will still be present today.

In runs 1-3, the number of smothered neutron stars produced greatly exceeds the number of LMXBs produced, by a factor $\sim 5$. Also, the number of white dwarf/white dwarf merger events produced via gravitational-radiation inspiral is approximately comparable to the production rate of LMXBs. One might therefore be optimistic that a large fraction of the MSPs observed in globular clusters may have been produced through evolutionary paths not including an LMXB phase, for example by accretion from the thick disk surrounding the neutron star in smothered systems (Krolik 1984). It should also be noted that the fate of a white dwarf/white dwarf merger is unclear; such an event may produce a supernova leaving no remnant, or the neutron produced may not have a low magnetic field or rotate rapidly (Phinney & Kulkarni 1994).

All three runs produced $\sim$ 300-400 blue stragglers.

Limiting the evolution time, and inserting the binaries at random times rather than at a time of zero both reduced the production of contact binaries and mergers via GR inspiral. From Table 10 we see that the number of LMXBs and CVs produced decreases from 121 to 48 comparing run 1 to run 3. We believe globular clusters are $\sim$ 15 Gyr old, and a steady flow of binaries into the core may be more physical than an instantaneous insertion of all of them at zero age. Hence, one might believe run 3 to be the physically more relevant of runs 1-3.

### 5.7 The Effect of Changing the IMF

In the first three runs, we considered the binary to be comprised of two $0.8393 M_\odot$ main-sequence stars. The actual binaries are more likely to be composed of stars having a range of masses, hence we now allow the binaries to be composed of stars of different mass, drawing the masses of the two stars independently from the IMF, with the additional criterion that the total mass be larger than $1 M_\odot$, this being typical of the average mass of stars in the cores of fairly concentrated clusters. In runs 4-6, we evolve the binaries for 15 Gyr, and inject them at random times as was done in run 3. We consider three values for the IMF, $\alpha$ having values of 2.35, 2.70, and 2.00 for runs 4, 5, and 6, respectively (where the IMF is given by the power law, $dN/dm \propto m^{-\alpha}$). The number of blue stragglers produced is seen to be a sensitive function of the IMF; fewer blue stragglers are produced for smaller values of $\alpha$. The number of smothered neutron stars also depends on $\alpha$. The flatter the IMF, the larger the number of neutron stars in the core, and thus the production rate of TZOs and LMXBs increases, whilst the production of CVs becomes less likely. This effect has already been noted by Hut & Verbunt (1984).

### 5.8 The Effect of the Evolution of Smothered Neutron Stars

We now consider the effect of assuming a smothered neutron star in a binary will expand to form a red giant and engulf its companion forming a CE phase, in a manner analogous to that for smothered white dwarfs, rather than evolve rapidly to some cataclysmic event such as a supernova explosion. In runs 7-10, we injected 1000 binaries at random times into the core of the cluster having $W_0 = 12.0$ and an IMF with $\alpha = 2.35$. In runs 8 and 9, the binaries always comprised of two $0.8393 M_\odot$ main-sequence stars, whereas in runs 7 and 10, the masses were drawn from the IMF, as described earlier. In runs 7 and 8 we evolved the binaries up to a time of 15 Gyr, whereas in runs 9 and 10 we evolved the binaries until one of the events described in §5.5 occurred. Comparing run 7 to run 4, we see only a slight increase in the number of LMXBs produced. There are 28 additional unresolved binaries. In other words, we have produced fewer objects containing neutron stars in smothered or contact systems in run 7 compared to run 4. It is also important to note that the ratio of LMXBs to smothered systems is also quite different between runs 4 and 7. In the former many more smothered systems are produced. Comparison between runs 7 and 10 show that if we are willing to wait long enough, we are eventually left with more neutron stars in contact or smothered systems. The number of neutron stars merging with white dwarfs is particularly enhanced, as is the number of white dwarf/white dwarf mergers. Comparison of runs 7 and 10, to 8 and 9, show that binaries containing more massive stars tend to lead to fewer unresolved binaries (comparing runs 7 and 8), and a boost in the production of blue stragglers.

### 5.9 The Effect of Changing $\alpha_{ce}$

In runs 11-13, we investigate the effects of changing the value of the efficiency of the common envelope stage, $\alpha_{ce}$. As mentioned earlier, it has been suggested that $\alpha_{ce} \sim 0.2 - 0.4$. In our previous runs, we have assumed $\alpha_{ce} = 0.4$, in runs 11, 12, and 13 we took $\alpha_{ce} = 0.2$, 0.1, and 0.05, respectively. In all three runs, we use the cluster model with $W_0 = 12.0$, and $\alpha = 2.35$. We inject the binaries at a random time between zero and 15 Gyr, and evolve the binaries up to the present time (15 Gyr). We treat smothered neutron stars in the same way as runs 7 - 10, *i.e.* assuming the envelopes expand to engulf the binary companion. The binaries are selected in the same way as run 4.

Decreasing the value of $\alpha_{ce}$ has the following effect: the number of unresolved binaries decreases, as CE phases leave the two stars much closer together. In the case of two compact objects they are sufficiently close to merge via gravitational radiation inspiral (hereafter denoted as MGRI) in more systems, hence we see a higher rate of such mergers. However, we note that the effect is relatively small: a factor of three between run 7 ($\alpha_{ce} = 0.4$) and run 13 ($\alpha_{ce} = 0.05$). The production rate of blue stragglers is seen to be independent of $\alpha_{ce}$.

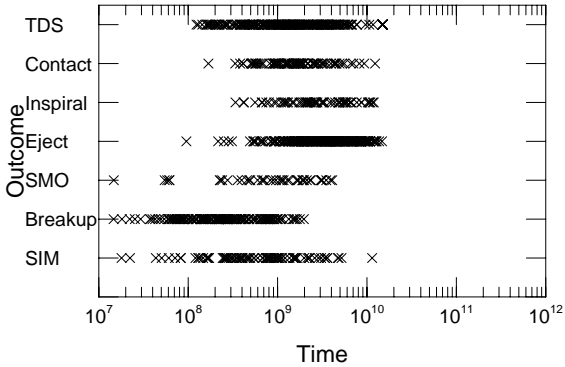

**Figure 8.** Final outcome of the binaries evolved in Run 17 (labelling same as Fig. 7).

### 5.10 Production Rates for a More-concentrated Cluster

We now consider encounters in a more concentrated cluster, where $W_0 = 16.0$. Run 14 uses the same parameters as run 13, except for the cluster model. Comparing these two runs, we note that the number of unresolved binaries is greatly reduced in run 14 as the encounter timescale is much shorter because of the denser cluster core. The number of MGRI events is also much larger in run 14. Owing to the higher concentration, the more collapsed cluster has much higher mass segration, and thus the core contains a larger fraction of neutron stars. It therefore comes as no surprise to see a larger fraction of the MGRIs containing neutron stars in run 14 than run 13. The number of blue stragglers produced in both runs are comparable.

In run 15, we repeat the conditions of run 14, except we use $\alpha_{ce} = 0.4$. This has the effect of reducing the number of MGRIs produced and slightly increasing the number of unresolved binaries, as two stars are left further apart after a CE phase.

Run 16 is a repeat of run 15, except the evolution is halted as soon as a smothered neutron star in produced (the outcome being labelled "TDS" as before). From 1000 binaries, some 220 smothered neutron stars are produced. Including the MGRI events, the ratio of smothered neutron stars to LMXBs is $\sim 10$ in run 16, compared to $\sim 3$ in run 15. Clearly deciding how smothered neutron stars evolve is a crucial question in computing the ratio of LMXBs to other systems containing neutron stars.

In run 17, we repeat run 16, except we inject all the binaries into the cluster core at a time of zero. This has the effect of greatly reducing the number of unresolved binaries. The number of ejected binaries increases, as does the number of smothered neutron stars. However, the number of MGRIs and contact binaries remains essentially the same as for run 16. In Figure 8 we plot the final outcomes of the binaries in run 17 as a function of time. Comparison with Figure 7 makes it clear that the evolutionary timescale for binaries in the more-concentrated cluster is significantly shorter, with all binaries reaching some end state in $< 10 \mathrm{Gyr}$.

Finally, in runs 18–20 we investigate the effect on production rates of changing the IMF. Runs 18 and 19 are repeats of run 16, except we use IMFs with $\alpha = 2.0$, and 2.7, respectively. As was the case for the cluster model with $W_0 = 12.0$, the number of blue stragglers produced is a sensitive function of $\alpha$; the number of blue stragglers being produced ranging from $\sim 100$ to $\sim 230$. The ratio of LMXBs to CVs is also seen to vary with $\alpha$; a larger $\alpha$ implies a smaller number of neutron stars relative to white dwarfs and thus the ratio of LMXBs to CVs is reduced. In run 20, we repeat run 18, but do not halt the evolution of the binary when a smothered neutron star is produced. Comparison of run 20 to run 18 shows a decrease of the ratio of smothered neutron stars to LMXBs.

### 5.11 Production Rates for Less-concentrated Clusters

We now consider less-contrated clusters. In runs 21-23, we take the conditions of run 16, except that we use three different models for less-concentrated clusters, having $W_0 = 8.0$, 6.0, and 4.0 respectively. In all three runs, the lower density of stars leads to a much longer timescale for encounters compared to the denser clusters considered earlier. The vast majority of the 1000 binaries remain unresolved. The number of blue stragglers produced is seen to decrease steadily with decreasing core density. Only a handful of contact binaries are produced in the three runs, and no MGRI events occurred.

### 5.12 Comparison of 2+1 to 1+1 Rates

We now compare the production rates of astrophysically interesting objects by 2+1 encounters to those from encounters between two single stars, listed for the various cluster models in Table 9. The encounter rates listed in Table 9 are given in units of encounters/$10^8$ years, hence assuming the globular clusters are 15Gyr old, we must multiply these numbers by 150 to obtain the total number produced (on our assuption that the population of single stars is static). In considering encounters between binaries and single stars, we injected 1000 binaries into the cluster cores. We must now consider how many binaries we expect to find in the cores of the various models. In the introduction, we stated that 2+1 encounters will be more important than 2+2 encounters provided that the mass fraction contained in binaries was small. Indeed in modelling the clusters as King-Mitchie models we neglected the effects of the binaries on the distribution of stars within the cluster. If we make the simple assumption that at a given time, $\sim 10\%$ of the stars in the core are contained in binaries, then we conclude from Table 8 that cluster models A-C contain $\sim 1000$-1500 binaries at any given time. For models D-F, the number is in the range 500-700, and for clusters G, H, and I, the numbers of binaries are 2500, 2000, and 1300 respectively. We now have to

**Table 11.** The total number of systems produced over 15 Gyr, via both single star/single star (1+1), and binary/single star (2+1) encounters.

| Model | BS | | NS | | | WD | | | MGRI | |
|---|---|---|---|---|---|---|---|---|---|---|
| | 1+1 | 2+1 | 1+1 | XB | SN | 1+1 | CV | SW | NW | WD$_a^2$ |
| A | 50 | 130 | 100 | 22 | 105 | 140 | 20 | 15 | 5 | 11 |
| B | 165 | 240 | 80 | 9 | 67 | 248 | 32 | 17 | 11 | 8 |
| C | 340 | 350 | 44 | 10 | 35 | 295 | 38 | 32 | 2 | 6 |
| D | 36 | 190 | 340 | 54 | 448 | 188 | 24 | 4 | 38 | 18 |
| E | 223 | 400 | 478 | 44 | 440 | 673 | 58 | 18 | 68 | 58 |
| F | 731 | 670 | 385 | 26 | 328 | 1278 | 98 | 66 | 70 | 110 |
| G | 35 | 160 | 14 | 2 | 4 | 41 | 4 | 14 | 0 | 0 |
| H | 18 | 80 | 1 | 0 | 0 | 9 | 0 | 0 | 0 | 0 |
| I | 7 | 20 | 0 | 0 | 0 | 3 | 1 | 0 | 0 | 0 |
| B′ | 165 | 220 | 80 | 17 | 2 | 248 | 24 | 11 | 31 | 35 |
| E′ | 223 | 374 | 478 | 48 | 4 | 673 | 62 | 18 | 276 | 84 |

BS is the number of blue stragglers produced. Under NS we list the number of neutron stars involved in 1+1 encounters, the number forming X-ray binaries (XB) via 2+1 encounters, and those in smothered systems formed via 2+1 encounters. We list the equivalent three numbers for white dwarfs under WD, with CV standing for cataclysmic variables. As in Table 10, systems brought into contact via gravitational radiation are labelled MGRI (mergers through gravitational radiation inspiral). Here we subdivide this outcome into: neutron star/white dwarf mergers (NW), and white dwarf/white dwarf mergers where the total mass is greater than the Chandrasekhar mass (WD$_a^2$). In models B′ and E′, we treat the evolution of a smothered neutron star in a binary in the same way as for white dwarfs and assume the envelope expands engulfing the other binary component and leading to a common envelope phase. We also take the common envelope efficiency $\alpha_{ce} = 0.05$ rather than the value used in the other runs listed ($\alpha_{ce} = 0.4$).

consider the average time taken by a binary injected into a core to reach one of the end states (*i.e.* ejection, merging *etc* ). For cluster models A-C, inspection of Table 10 (runs 4-6) reveals that ∼ 500 of the original 1000 binaries have reached some end state in 15 Gyr. For models D-F, this number is ∼ 0 − 200. In fact, in these more collapsed clusters, ∼ 50% of the binaries have reached some end state in ∼ 1 − 2 Gyr. Hence for cluster models D-F, we might imagine that in 15 Gyr, the core is able to "process" ∼ 2000 − 4000 binaries. In all other cases, our assumptions here seem to suggest that the cores will be unable to deal with more than 1000 binaries. In the case of models G-I, only a very small fraction of injected binaries will reach some end state, as seen in Table 10. In other words when comparing the rates for encounters between two single stars given in Table 8 to those for encounters between single stars and binaries given in Table 10, we should apply the following factors to the latter: ∼ 1 for models A-C, ∼ 2 − 4 for models D-F, and ∼ 1 − 2 for models G-I. Assuming encounters happen at a uniform rate over 15 Gyr, we should then multiply the rates in Table 9 by a factor of 150.

In Table 11, we list the total expected production of blue stragglers and systems containing compact objects, assuming the adjustment factors discussed in the previous paragraph. Considering first the production of blue stragglers, we see that the ratio of the number produced via 2+1 encounters to the number produced via 1+1 encounters is a sensitive function of cluster concentration and IMF. In models A-C, roughly equal numbers of blue stragglers are produced through both pathways. We note again that a larger value of the IMF power law, $\alpha$, implies a greater mass fraction contained in main-sequence stars, which leads to an increase in the production of blue stragglers. We now consider the most concentrated clusters, models D-F. In model D, we produce very few blue stragglers from 1+1 encounters. We have a relatively flat power-law IMF, and the lower-mass main-sequence stars have been largely excluded from the core by the effects of mass segregation. In models E and F, the steeper power-law IMF, means more main-sequence stars remain in the core, and the number of blue stragglers produced are larger than for models B and C because of the increased density. By comparison, the number of blue stragglers produced via 2+1 encounters show less variation between models D-F. In models D and E, the number of 2+1 events is greater than the number of 1+1 events, whereas for model F, the two numbers are comparable. Finally, looking at the least concentrated clusters considered, namely models G-I, we see that 2+1 encounters are significantly more important than 1+1 encounters in producing BSs in low-density clusters. We also note that in models B′ and E′ in Table 11, we have changed the way we deal with smothered neutron stars contained in binaries, treating them exactly the same as white dwarfs. One would expect this to have little effect on the production of BSs, an expectation seen in the results.

We next consider the number of systems produced con-taining neutron stars. Looking first at models A-C, we see that the number of neutron stars involved in 1+1 encounters is roughly equal to those in X-ray binaries or smothered systems produced from 2+1 encounters. Both numbers show a strong dependence on the IMF power law, $\alpha$, with a flatter IMF leading to an increase in the number of systems produced. We note also that the number of systems brought into contact via gravitational radiation is relatively small. If we assume systems produced via 1+1 encounters all form smothered systems, then our ratio of smothered systems to X-ray binaries $\sim$ 10:1. Considering next the most-concentrated clusters, models D-F, we see that the number of systems produced has increased compared to models A-C owing to the greater densities and increased core mass fraction contained in neutron stars due to stronger mass segregation effects. We also note that the number of MGRI systems has increased; owing to the shorter encounter timescale in the densest cores, systems pass through a common envelope stage at an earlier time, and thus have more time to spiral together. If we include all neutron star/white dwarf and white dwarf/white dwarf (with total mass above the Chandrasekhar limit) coalescences in the total number of smothered neutron star systems, then this number will be enhanced by 10–60 % for models D-F. With the same assumptions as above, we expect the ratio of smothered systems to X-ray binaries $\sim$ 15:1 for model D, and $\sim$ 30 : 1 for model F. We thus see that it is quite possible that a much larger number of systems are produced containing smothered neutron stars. If these systems evolve into MSPs, they may do so without passing through a prolonged phase of X-ray emission, hence we would not expect to see roughly equal numbers of LMXBs and MSPs in globular clusters. Assuming that LMXBs have a lifetime $\sim$ 1Gyr, then we would expect to see $\sim 1 - 3$ LMXBs in cluster models A-F. Let us now consider models B$'$ and E$'$. As mentioned earlier, in these runs we treated smothered neutron stars in binaries in a similar way to smothered white dwarfs, assuming their gaseous envelopes expanded to engulf the other component of the binary to enter a common envelope phase. In order to maximise the number of MGRI products, we considered a relatively low value for the common envelope efficiency, $\alpha_{ce} = 0.05$, which will leave post-CE phase stars closer together than larger values. We note that the number of X-ray binaries is increased slightly (comparing B$'$ to B, and E$'$ to E). The total number of smothered systems produced via 2+1 encounters is reduced only slightly though this result is very dependent on the value of $\alpha_{ce}$. The ratio of smothered systems to X-ray binaries $\sim$ 10 : 1 for model B$'$, and $\sim$ 20 : 1 for model E$'$, these ratios being reduced by a factor of two when $\alpha_{ce}$ is increased to 0.4. Hence we see the importance of determining the subsequent evolution of smothered neutron stars, and calculating the value of $\alpha_{ce}$.

We should remind ourselves of the uncertainty in the outcomes of encounters between a neutron star and a main-sequence star. In the ratios computed above, we have assumed that all such encounters will produce smothered neutron stars. In previous work (Davies, Benz & Hills [1992]), we showed that at least $\sim$ 0.5 of the encounters between single main-sequence and neutron stars would produce smothered systems. If only one half actually did so, then the ratio of smothered systems to X-ray binaries $\sim$ 5 : 1 for models A-C, with similar reductions in the other models. Hence we conclude two things: 1) the fraction of single encounters producing clean binaries would be a useful number to determine, and 2) even if half such encounters produce clean binaries we still seem likely to produce a much larger number of smothered systems, the pessimistic factor of five mentioned here possibly representing a useful step towards explaining the excess of MSPs compared to LMXBs as seen in globular clusters today.

Finally, we consider the number of systems produced containing white dwarfs. In some sense, comparison of the 1+1 to the 2+1 rates is misleading, as we have assumed in all 2+1 runs that a smothered white dwarf in a binary will essentially form a red giant and engulf the other companion to form a common envelope system. If we assume that none of the systems produced via 1+1 encounters will produced CVs, then we conclude that the number of CVs will be very much smaller than the number of smothered white dwarfs produced for models A-F. We expect the smothered white dwarfs to evolve to form red giants; the compact object accreting material as the envelope expands. In the case of the most-massive white dwarfs smothered by the remains of a main-sequence star, it is possible that it will accrete enough material to exceed the Chandrasekhar mass, possibly forming a neutron star via AIC. The number of white dwarf/white dwarf mergers occuring via gravitational radiation inspiral where the total mass is *less* than the Chandrasekhar limit is negligible for all models.

## ACKNOWLEDGEMENTS

We thank Sverre Aarseth, Willy Benz and Steinn Sigurdsson for useful discussions, and the anonymous referee for their comments. The support of an R.C. Tolman Research Fellowship from Caltech is gratefully acknowledged.

## REFERENCES

Binney J. J., Tremaine S., 1987, in Galactic Dynamics. Princeton, p.235
Cannon R. C., 1993, MNRAS, 263, 817
Cleary P. W., Monaghan J. J., 1990, ApJ, 349, 150
Davies M. B., Benz W., Hills J. G., 1992, ApJ, 401, 246
Davies M. B., Benz W., Hills J. G., 1993, ApJ, 411, 285
Davies M. B., Benz W., Hills J. G., 1994, ApJ, 424, 870
DiStefano R., Rappaport S. A., 1992, ApJ, 396, 587
Grabhorn R. P., Cohn H. N., Lugger P. M., Murphy B. W., 1992, ApJ, 392, 86
Hut P., McMillan S., Romani R. W., 1992, ApJ, 389, 527
Hut P., Verbunt F., 1984, Nature, 301, 587
Iben I. Jr., Renzini A., 1983, ARA&A, 21, 271
Kochanek C. S., 1992, ApJ, 385, 604
Krolik J., 1984, ApJ, 282, 452
Lightman A. P., Shapiro S. L., 1977, ApJ, 211, 244
Meylan G., 1987, A&A, 184, 144
Meylan G., 1989, A&A, 214, 106
Phinney E. S. P., Kulkarni S., 1994, ARA&A, 32, 591
Pryor N., Meylan G., 1993, in Djorgovski S. G., Meylan G., eds, ASP Conference Series vol. 50: Structure and Dynamics of Globular Clusters. p.357
Sigurdsson S., Phinney E. S. P., 1994, ApJ, submitted
Taam R. E., Bodenheimer P., 1989, ApJ, 337, 849
Verbunt F., Zwaan C., 1981, a&a, 100, L9